\documentclass[english,prl,onecolumn]{revtex4}
\usepackage[T1]{fontenc}

\usepackage{subfig}
\usepackage{pdfpages}
\usepackage{epstopdf}
\usepackage{babel}
\usepackage{amssymb,amsmath}
\usepackage{blindtext}
\usepackage{lipsum}
\usepackage{MnSymbol}

\begin{document}

\title{Micro-foundation using percolation theory of the finite-time singular behavior of 
the crash hazard rate in a class of rational expectation bubbles}

\author{Maximilian Seyrich$^{1}$$^,$$^{2}$ and Didier Sornette$^{3}$$^,$$^{4}$}

\affiliation{$^1$ Department of Physics, ETH Zurich, Otto-Stern-Weg 1, 8093 Zurich, Switzerland, mseyrich@ethz.ch\\$^2$ Institut f\"ur Theoretische Physik, Technische Universit\"at Berlin, Hardenbergstrasse 36, 10623 Berlin, Germany\\
$^3$Chair of Entrepreneurial Risk, ETH Zurich, Scheuchzerstrasse 7, 8092 Zuerich, Switzerland, \\
$^4$ Swiss Finance Institute, c/o University of Geneva, 40 blvd. Du Pont d'Arve, CH 1211 Geneva 4, Switzerland}

\date{\today}

\begin{abstract}
We present a plausible micro-founded model for the previously postulated
power law finite time singular form of the crash hazard rate in the Johansen-Ledoit-Sornette model 
of rational expectation bubbles. The model is based on a percolation picture of the network of traders and the concept that clusters of connected traders share the same opinion. The key ingredient is the notion that a shift of position from buyer to seller of a sufficiently large group of traders can trigger a crash.
This provides a formula to estimate the crash hazard rate by summation over percolation clusters
above a minimum size of a power $s^a$ (with $a>1$) of the cluster sizes $s$, similarly to a generalized percolation susceptibility.
The power $s^a$ of cluster sizes emerges from the super-linear dependence of group activity as a function
of group size, previously documented in the literature. The crash hazard rate exhibits
explosive finite-time singular behaviors when the control parameter (fraction of occupied sites, or density
of traders in the network) approaches the percolation threshold $p_c$. Realistic dynamics
are generated by modelling the density of traders on the percolation network by an Ornstein-Uhlenbeck 
process, whose memory controls the spontaneous excursion of the control parameter
close to the critical region of bubble formation. Our numerical simulations recover the main
stylized properties of the JLS model with intermittent explosive super-exponential bubbles interrupted by crashes. 
\end{abstract}

\maketitle

\section{Introduction}

Many financial crashes, including the crash of October 1929 followed by the Great Depression or the recent financial crash that started in 2007 followed by the Great Recession, were preceded by at least one massive bubble \cite{sornetterigoros,Sorcauill14}. 
Most policy makers and regulators, as well as most academics, have been unable to detect these bubbles ex-ante. 
Only after their bursts occurred
were their presences confirmed and accepted in the general wisdom. One fundamental reason for the difficulty in identifying bubbles ex-ante
lies in the standard definition of a bubble, which is supposed to be a deviation of the observed price from the fundamental value.
Since the fundamental value is notoriously difficult to evaluate in practice (cf the famous factor of two of Fisher Black \cite{Black86}),
reliable diagnostics of deviations from a quantity that is very noisy to start with are bound to fail or be very imprecise.

Since \cite{SorJohBou96,FeigenbaumFreund96}, another approach has been proposed that consists in characterising financial bubbles
as transient super-exponential dynamics decorated by accelerating volatility patterns. The term "super-exponential"
refers to the property of prices to grow transiently faster than a (noisy) exponential, i.e., the growth rate grows itself
during bubble regimes. Recall that, in contrast, a standard exponential growth, the average trajectory of the Geometrical Brownian Motion and of most of the financial models of prices, corresponds to a constant average growth rate.
A number of recent experimental \cite{HueslerSorHom13} and empirical investigations 
\cite{Jiangchin10,sornetterigoros,LeissNaxSor15,SorCaubub15} have reinforced the credibility of this 
property characterizing bubbles. 

Such super-exponential
process was made operational in the Johanson-Ledoit-Sornette (JLS) model of bubbles and crashes \cite{JLS,JLS2,Johansen1999,vladimir}.
The model considers a single asset that is purely speculative and pays no dividends in a market where we ignore the interest rate, and use simplified market-clearing conditions. A number of extensions have been developed in 
\cite{ZhouSorfund06,SorZhoufor06,YanRedaetal12,Yanwoodsor12,Sorclarif13,LinSor13,YanWoodSorinf14,LinRenSor14}.
Embedded in the general framework of rational expectation (RE) bubbles, the JLS model views the stock market as
divided into investors with rational expectations and noise traders. The latter are influenced by other traders and thus may exhibit herding behavior.  Local self-reinforcement of the decisions of noise traders and the resulting herding effects are responsible for the bubble as well as the crash. A crash may occur if the strength of the herding effects exceeds a critical value and many traders sell at the same time. 
For motivations of these herding effects, see Refs. \cite{CMC,Sornette2004}. 

Due to the ubiquity of noise, the JLS model proposes a stochastic description for the occurrence
of a crash in terms of a crash hazard rate.  By definition, the crash hazard rate (multiplied by $dt$) is the conditional probability 
that the crash occurs in the time interval $[t,t+dt]$, given that it has not yet happened until time $t$.
In the RE JLS model, a crash is not a deterministic outcome of the bubble.
Instead, the time of the crash, if it happens, is a random variable. Moreover, the bubble may end without a crash.
Therefore, it is rational for traders to stay invested in the asset as long as the return remunerates for the risk of the crash. 

In the JLS model \cite{JLS,JLS2,Johansen1999}, the crash hazard rate is postulated to exhibit a 
power law finite time singular behavior, decorated by log-periodic structures. Here, we just present the first
component, the power law finite time singularity in the form
\begin{equation}
h(t) \propto {1 \over (t_c-t)^{\alpha}}~,
\label{rwthryt}
\end{equation}
where $t_c$ is the end of the bubble, corresponding to the time when the crash is most probable (but not certain), and $\alpha >0$ quantifies the strength of the singularity.
This form was motivated on the basis
of an analogy with critical phenomena, in which imitative Ising-like agents exhibit global coordinated behaviors when the forces of imitation increase to overcome the amplitude of randomness (or idiosyncratic opinions). See \cite{SorZhouIsing06,ZhouSIsingor07,HarrasSor11,HarrasTessSor12} for specific implementation of Ising-like traders. A large shift in the market book, leading to a crash according to the Kyle model \cite{Kyle}, is possible in this herding phase. 
By the RE condition, an increasing crash hazard rate translates into a correspondingly accelerating
instantaneous return that is necessary for the traders to remain invested, given the rising crash risk.
The JLS model thus uses a risk-driven price mechanism, where the dynamics of the crash hazard rate
governs that of the price. 

The purpose of the present article is to present a plausible micro-founded model for the previously postulated
power law finite time singular form of the crash hazard rate.

\section{Brief summary of the mathematical formulation of the JLS model \label{ynhtbfw}}
 
The dynamics of the price evolution is given by
\begin{equation}
\label{eq:price}
\frac{dp_{\textrm{rice}}(t)}{p_{\textrm{rice}}}=  \rho(t) dt + \eta dw(t) - \kappa dj(t)~,
\end{equation}
where $\rho(t)$ is the instantaneous return (or growth rate of the price) that remunerates the investors
for being exposed to the market risks. There are two types of risks.
The first risk is the standard volatility risk 
represented by the second term $\eta dw(t)$ in the r.h.s. of expression (\ref{eq:price}),
where $\eta$ is the volatility and $dw$ is the standard increment of the Wiener process (which has
zero mean and variance equal to $dt$). The second risk is the possible occurrence of crashes,
represented by the last term  $- \kappa dj(t)$ in the r.h.s. of expression (\ref{eq:price}).
The coefficient $\kappa$ is the amplitude of a crash when it occurs. The process $j(t)$ is
a counting process that jumps by $1$ each time there is a crash. Thus, $dj(t)$ is equal to $0$ at all times,
except when a crash occurs, at which time it takes the value $1$. The time of the crash $t^*$ is a random variable, which is described by the a priori deterministic crash hazard rate $h(t)$. However, the crash hazard rate itself can have a dependency on another stochastic process, too. We will later implement this when our control parameter is described by an Ornstein-Uhlenbeck process. The crash hazard rate is defined such that the probability that a crash occurs in the time interval $t$ and $t+dt$ is given by $h(t)dt$. 

Taking the expectation of expression (\ref{eq:price}) yields
\begin{equation}
\label{eq:priceex}
{\rm E}\left[\frac{dp_{\textrm{rice}}(t)}{p_{\textrm{rice}}}\right]=  \rho(t) dt - \kappa {\rm E}\left[dj(t)\right] =  \rho(t) dt - \kappa  h(t) dt~.
\end{equation}
The last equality results from the evaluation of 
${\rm E}\left[dj(t)\right]$, which is given by ${\rm E}\left[dj(t)\right] = 1 \times h(t) dt + 0 \times (1- h(t) dt) = h(t) dt$.
For risk adverse traders, ${\rm E}\left[\frac{dp_{\textrm{rice}}(t)}{p_{\textrm{rice}}}\right]$ should be equal to 
the risk premium that they require to remain invested in the market. To simplify notation and without changing
the results, it is convenient to assume that traders are risk-neutral, which amounts to stating that ${\rm E}\left[\frac{dp_{\textrm{rice}}(t)}{p_{\textrm{rice}}}\right]=0$. Replacing $0$ by a constant non-zero value just amounts to adding a drift in the price process
without essential modification of the dynamics. The condition of vanishing of the r.h.s. of expression (\ref{eq:priceex})
leads to the remarkable generalisation of the return-risk relationship in the presence of crashes, namely
\begin{equation}
\rho(t) = \kappa h(t)~,
\label{wrhtwyrtbq}
\end{equation}
which is the pillar of the JLS model and its extensions \cite{ZhouSorfund06,SorZhoufor06,YanRedaetal12,Yanwoodsor12,Sorclarif13,LinSor13,YanWoodSorinf14,LinRenSor14}.
Replacing (\ref{wrhtwyrtbq}) in (\ref{eq:price}) gives the dynamics of the price
\begin{equation}
\label{eq:pric2t4e}
\frac{dp_{\textrm{rice}}(t)}{p_{\textrm{rice}}}=  \kappa h(t) dt + \eta dw(t) - \kappa dj(t)~.
\end{equation}
Expression (\ref{eq:pric2t4e}) demonstrates that the price dynamics is essentially specified once
the functional form of the crash hazard rate $h(t)$ is known, as it determines both the drift
and the jump process $j(t)$, up to the standard Brownian motion component $\eta dw(t)$.
The remaining focus of the present article is the development of a micro-founded model for $h(t)$.
 
Because our numerical simulations will be using discrete times, it is convenient to reformulate 
expression (\ref{eq:pric2t4e}). For this, we replace the continuous time Wiener process
by a discrete time symmetric random walk, whose increment is denoted $\Delta w$ in the following. The symmetrical random walk is the pendant of the Wiener process as it converges to a Wiener process in the limit of an infinite number of time-steps \cite{donsker}.
The price process in discrete times is given by
\begin{equation}
\label{eq:pricesim}
p_{\textrm{rice}}(t)= p_{\textrm{rice}}(t-1) \left[\kappa h(t) \Delta t - \kappa \Delta j(t) + \eta \Delta w(t) \right] ~,
\end{equation}
where the time step is $\Delta t$ and $\Delta j(t)=0$ except at the time of a crash when it is equal to $1$.

\section{Micro-founded model of the Crash hazard rate}

\subsection{Model formulation}

The model that we now present is reminiscent of the percolation-based model of 
Cont and Bouchaud, in which clusters of agents with similar opinion control the size of the trading positions
\cite{contbouchaud}. The dynamics of the price then derives from that of the cluster distribution.
Similarly to \cite{contbouchaud}, the universe of traders is partitioned in clusters,
such that each cluster is characterized by a well-defined single opinion, via a process of homophily.
Clusters represent for instance mutual funds or the result of herding among security analysts evaluating a stock market. 

We use graph theory to describe the network of interactions between traders. Traders sit at nodes of
a graph and a link exists between two traders when they can influence each other via their physical 
or digital proximity. The whole market is constructed as the graph of all connections between pairs of traders.
The overall strength of imitation is embodied in a single control parameter $p$, which represents
the average fraction of existing links or nodes. As $p$ is varied from $0$ to $1$, percolation theory predicts
the existence of a well-defined critical value $p_c$ at which a transition occurs between
a graph made of finite disconnected clusters to a graph containing an infinite cluster coexisting
with finite clusters \cite{Grimmett89,Grimmett97,Stauff}. This is the so-called percolation transition.

In percolation theory \cite{Grimmett89,Grimmett97,Stauff}, a crucial quantity is the distribution of cluster sizes.
Let us denote by $s$ the size of a cluster of connected traders. Then, $n_t(s)$ is defined as the number of clusters
of size $s$ per node at time $t$. This definition, which is standard in percolation theory, implies
that the total fraction of nodes belonging to all clusters of size $s$ is $s n_t(s)$. Then, the sum over
all sizes, $\sum_{s=1}^\infty s n_t(s)$, must be normalized to $1$, to represent the fact that the sum
of all node fractions over all cluster sizes exhausts the full enumeration of all sites within the graph.
Note that we put a subscript $t$ in $n_t(s)$ to allow for a time dynamics in the graph
topology that results from continuous link formations and removals.

In our model, each cluster contains a collection of connected traders who are similarly minded, such that 
all persons within a cluster adopt the same market position.
At a given time, we can enumerate the positions $\pi$ (`long' or `short') of each cluster in the system:
$\{\pi[s_t^{(1)}], \pi[s_t^{(2)}], ..., \pi[s_t^{(N)}]\}$. Representing a long position by $+1$ and a short position by $-1$,
the position $s_t^{(i)} \pi[s_t^{(i)}]$ of cluster $i$ is thus either $+s_t^{(i)}$ or $-s_t^{(i)}$. This means that the 
number of shares bought or sold by the $s_t^{(i)}$ traders in that cluster is simply proportional to 
the number of traders. The net supply-demand is
thus the sum $\sum_{i=1}^N s_t^{(i)} \pi[s_t^{(i)}]$. It happens that, as a result of novel pieces of information
or due to other idiosyncratic processes, the opinions of traders within some clusters change from
$\pi[s_t^{(i)}]=+1$ to $\pi[s_{t+1}^{(i)}]=-1$, corresponding for instance to change from a long to a short position
(this amounts to selling $2 s_t^{(i)}$ shares). Note that we can replace the short position to neutral (holding
only cash and preventing any short position) without change of the main structure of the model.
In the standard double-auction mechanism of price fixing,
the price moves so as to maximize the number of transactions, i.e., in such a way as to satisfy the requests of
the maximum number of traders.

The novel ingredient of our model is to assume that a financial crash occurs when a cluster of sufficiently large size
changes its position from long (bullish) to short (bearish). For simplicity, we assume that there exists a
size threshold $s_m$ such that, when a cluster of size larger than or equal to $s_m$ shifts from bullish to bearish,
the corresponding market impact triggers a domino effect of selling of other clusters, leading to a market crash.
The motivation for such a behavior is, for instance, found in the mechanism of portfolio insurance of Leland and
Rubinstein \cite{LelandRubi76} that was determined to have played a key role in aggravating the crash of Oct. 19, 1987
\cite{LelandRuaftercrash88,BradyReport88}. More generally, we invoke the set of mechanisms 
at the origin of positive feedbacks at work in financial markets and especially during times of 
large volatility spikes \cite{SorCaubub15}. Pro-cyclical increase of margin requirement, drying of liquidity
and panic are among the mechanisms that can lead to exacerbate price moves, amplifying them up 
to crash proportion. They act in a non-linear fashion such that a trigger of sufficient amplitude
can grow to enormous proportion, leading to a coordinated sell-off of a large number of agents, namely a crash.

Let us consider a given cluster $i$ of size $s_i$, which has at time $t$ the position $+s_t^{(i)}$.
We need to specify what is the probability that such a cluster changes its position to $-s_t^{(i)}$.
We assume that, as soon as one trader of the group shifts her opinion, the cohesion within the
cluster is strong enough to trigger a collective shift of all agents of that cluster in the same direction.

The change of a whole cluster position is thus determined by the change of at least one person
in that group. We argue that this change can result from two classes of mechanisms, 
{\it interaction-based} and {\it large deviation}. In other words, a trader in the group can change her opinion as a result
of her repeated interactions with other traders. Or she can be ``the one'' who is first exposed
to a novel information, or who just shifts her psychology as a result of numerous conscious or unspecified unconscious reasons.
In \cite{SaichevSor14,suplinOSS}, it was shown that, close to a regime of criticality, the activities or productivity
of a group of size $s$ has a super-linear dependence on the group size. As mentioned, this may result
from the appearance of a cooperative process and order involving a collective mode of 
traders defined by the build up of correlation between their contributions. Or, viewing the activity 
of opinions and their changes as a point process, the super-linear activity as a function of group size
can be shown to result from the sampling of an underlying heavy-tailed distribution of pre-existing
individual activities \cite{SaichevSor14}. In other words, if people are very different in their level of occupation and functioning,
the larger the group size, the larger is the probability to encounter an agent who has a very large activity
and this agent will be likely the one triggering the shift of opinion in the cluster.
There is also empirical evidence that social variables (such as productions or new inventions) of cities or more
generally of human groups scale super-linearly with their population \cite{Bettencourtetal07,suplincity,Schlapferetal14}. 
We thus postulate that the probability $R(s)$ per unit time with which a given cluster of size $s$ may change its position 
from bullish to bearish is given by
\begin{equation}
R(s) \propto s^a~,~~~~~~{\rm with}~ a >1~,
\label{hnenthwg}
\end{equation} 
where the condition $a>1$ embodies the super-linear relationship discussed above.

Putting all the above considerations together and using the notation of expression (\ref{eq:pricesim}),
the crash hazard rate can be obtained as
\begin{equation}
\label{eq:crashhaza}
h(t) \Delta t= \sum \limits_{s=s_m}^\infty [R(s) \Delta t] n_t(s) ~.
\end{equation}
This equation (\ref{eq:crashhaza}) expresses that the probability $h(t) \Delta t$ of a crash 
occurring at time $t$, given that it has not yet occurred, is proportional to the sum of the probabilities
that an agent of a cluster $s$ changes her opinion, where the sum runs
over all clusters larger than or equal to the threshold size $s_m$ to ensure sufficient impact.
As given by (\ref{hnenthwg}), $R(s) \Delta t$ is the probability that at least one agent actively changes
her opinion within a given cluster of size $s$ during the time interval $\Delta t$. 
This is weighted by the normalized number $n_t(s)$ of such clusters of size $s$,
assuming that any shift in a given cluster may occur independently of all the other clusters of the same size.
Then, the independence of these shifts across all types of clusters translates into the summation in the r.h.s.
of expression (\ref{eq:crashhaza}).

\subsection{Leading power law dependence of the crash hazard rate}

We can estimate the leading scaling behavior of $h(t)$ from the following well-known scaling relations of percolation theory \cite{Grimmett89,Grimmett97,Stauff}. By assuming that the control parameter $p(t)$ controlling the connectivity of percolation clusters is a smooth function of time, we obtain to $h(t) = h(p(t))$. First, the distribution of cluster sizes is given by
\begin{equation}
n(s) \sim {1 \over s^{1+\mu}} ~f(s/s^*)~,
\end{equation}
where
\begin{equation}
\label{eq:sstar}
s^* \propto \frac{1}{|p-p_c|^{\frac{1}{\sigma}}}~,
\end{equation}
and $f(\cdot)$ is a function decaying rapidly to $0$ as its argument becomes larger than $1$.
This implies that the sum in the r.h.s. of (\ref{eq:crashhaza}) can be truncated at the upper bound $s^*$, such that
\begin{equation}
\label{eq:crashefbte1haza}
h(t) \Delta t \approx \sum \limits_{s=s_m}^{s^*} [R(s) \Delta t] n_t(s) ~.
\end{equation}
Using expression (\ref{hnenthwg}) in (\ref{eq:crashefbte1haza}), we find the leading scaling behavior
of the crash hazard rate
\begin{equation}
h(p(t))\simeq \frac{-1}{\mu - a} \left. \frac{1}{s^{\mu -a}} \right|_{s=s_m}^{s=s^*(p(t))} =
\begin{cases}
\frac{1}{\mu - a} \left(\frac{1}{s_m^{\mu -a}} - \frac{1}{{(s^*)}^{\mu -a}}\right) &\text{ for $a < \mu$,} \\ \frac{1}{a-\mu} \left({s^*(p(t))} ^{a-\mu}- s_m ^{a-\mu}\right) &\text{ for  $a > \mu$.}
\end{cases}
\end{equation}
We are interested in the regime close to percolation criticality where $s^*(p) \gg s_m$. This leads to the simplification
\begin{equation}
h(p) \simeq
\begin{cases} \frac{1}{\mu - a} \frac{1}{s_m^{\mu -a}} &\text{ for $a < \mu$,} \\ \frac{1}{a-\mu} {s^*(p(t))} ^{a-\mu} &\text{ for  $a > \mu$.}
\end{cases}.
\label{wrynrtbq}
\end{equation}

Percolation theory predicts that the exponent $\mu$ of the distribution of cluster sizes is given by
\begin{equation}
\mu = {d \over d-{\beta \over \nu}}~,
\label{wytjuehw}
\end{equation}
where $d$ is the dimension of space, $\beta$ is the exponent of the order parameter and $\nu$ is the
exponent of the correlation length of percolation systems. In two-dimensional percolation ($d=2$), 
the exact values are known: $\beta = 5/36$ and $\nu = 4/3$, leading to $\mu_2 = 96/91 \approx 1.055$.
In three-dimensional percolation ($d=3$), $\beta \approx 0.418$ and $\nu \approx 0.876$, leading to $\mu_3 \approx 1.189$.
In dimensions $d \geq 6$, $\beta=1$ and $\nu = 1/2$ and $\mu_{d \geq 6} = 1.5$.
Thus, the exponent $\mu$ varies from slightly larger than $1$ in two dimensions to $1.5$ in arbitrary large dimensions.
Note that the case of large dimensions is perfectly possible when representing the network of social connections between
individuals, even if they live in a three dimensional space.

The behavior of $h(t)$ given by (\ref{wrynrtbq}) depends on the value of the exponent $a$ in (\ref{hnenthwg})
relative to $\mu$. For the production of large groups, such as cities,  Refs. \cite{Bettencourtetal07,suplincity,Schlapferetal14}
report exponents $a$ in the range from $1.1$ to $1.2$. For smaller groups involved in creative activities,
Ref. \cite{suplinOSS} documents exponents $a$ ranging from $1$ to more than $3$, with a median value $\approx 4/3$,
thus typically much larger than in  \cite{Bettencourtetal07,suplincity,Schlapferetal14}.
Ref. \cite{suplinOSS} also provides a reconciliation of these two set of values by recognising that the activity
of a large social group is the result of sub-groups, whose total contributions lead to a renormalisation
to a smaller exponent $a$. We conclude that the vast majority of groups reported in \cite{suplinOSS}
live in the regime $a>\mu$ for networks in two and three dimensions, while approximately 40\% of the groups
have their activity exponent $a>1.5$, so that the regime $a>\mu$ is verified in all dimensions.

\subsection{Generalisation accounting for the variability of the exponents describing super-linear activity}

Having argued that the regime $a>\mu$ is likely to describe the activity of many social groups, there is
an additional reason for the relevance of this regime.
Given the variability of the observed exponent $a$ mentioned above, it is sufficient that a not-too-small
fraction of the groups have $a>\mu$ for the second regime in (\ref{wrynrtbq}) to be relevant
in the total population of clusters. To see this, let us assume that the exponent $a$ is not unique
but is instead distributed according to the probability density function (pdf) $Q(a)$. What we mean by this
is that, taking a cluster at random, the activity $R_a(s)$ of that cluster of size $s$ is given by expression (\ref{hnenthwg})
with an exponent $a$ drawn from the pdf $Q(a)$. Then, expression (\ref{eq:crashefbte1haza}) is replaced by
\begin{equation}
\label{eq:crashefbteqreg1haza}
h(t) \Delta t \approx  \sum \limits_{s=s_m}^{s^*(t)} \int_0^\infty da Q(a) [R_a(s) \Delta t] n_t(s) ~.
\end{equation}
Replacing the  different terms by their power law relations, we obtain
\begin{equation}
\label{eq:crasyr2hefbteqreg1haza}
h(t) \approx  \sum \limits_{s=s_m}^{s^*(t)} \int_0^\infty da Q(a) s^a s^{-1-\mu} = \int_0^\infty da {Q(a) \over a -\mu}
\left[(s^*(t))^{a-\mu} - (s_m)^{a-\mu} \right]~,
\end{equation}
where the last expression in the r.h.s. is obtained by commuting the sum over $s$ and the integral over $a$
and performing the integration over $s$. In the critical regime of interest where $s^* \gg s_m$,
the integral over $a$ is completely dominated by the values of $a$ larger than $\mu$. This amounts to 
approximate expression (\ref{eq:crasyr2hefbteqreg1haza}) by 
\begin{equation}
\label{rhaaza}
h(t)  \approx  \int_\mu^\infty da {Q(a) \over a -\mu} (s^*(t))^{a-\mu} ~,
\end{equation}
where we have omitted the negligible term $(s_m)^{a-\mu}$.
Making the change of function $Q(a) := e^{q(a)}$ without loss of generality, we rewrite expression (\ref{rhaaza}) as
\begin{equation}
\label{rhaafqeqza}
h(t)  \approx  \int_\mu^\infty da ~e^{q(a) -\ln(a -\mu) + (a-\mu) \ln s^*(t)} ~.
\end{equation}
The integral can be evaluated by the saddle-node method, which yields
\begin{equation}
h(t)  \approx  (s^*(t))^\xi~,
\label{wrthyju}
\end{equation}
where the exponent $\xi$ is the solution of 
\begin{equation}
{dq \over da}|_{a=\xi} - {1 \over \xi -\mu} +  \ln s^*(t)  = 0~.
\label{mumnhgs}
\end{equation}
Since the function $q(a)$ is a smooth function of $a$, in the limit of interest where $s^* \to +\infty$,
the solution of (\ref{mumnhgs}) is asymptotically
\begin{equation}
\xi = \mu + {1 \over \ln s^*}  \to \mu  ~~{\rm for}~ s^* \to +\infty~.
\label{tumiujethw}
\end{equation}
Given the behavior of $s^*$ described by (\ref{eq:sstar}), 
this reasoning demonstrates the robust nature of the singular behavior of the crash hazard rate
\begin{equation}
h(t)  \approx  (s^*(t))^\xi \sim \frac{1}{|p(t)-p_c|^{\frac{\xi}{\sigma}}} \approx  \frac{1}{|p(t)-p_c|^{\frac{\mu}{\sigma}}}~,
\label{wrthwhyfdqfdju}
\end{equation}
which is valid as soon as some groups exhibit a sufficiently strong super-linear activity in their change of opinions
as a function of their group size. From the known scaling relations of percolation theory, 
we have the identity $\mu / \sigma = \nu d$, which gives a more straightforward evaluation for the exponent
controlling the leading behavior of the crash hazard rate as a function of the distance to the critical 
percolation point.

\vskip 0.3cm
\noindent
\textbf{Main Result:} {\it
Assuming that  the fraction $p(t)$ of existing links in the network of traders increases linearly with time, 
it will reach the critical percolation value $p_c$ at some time $t_c$. Then, close to $t_c$, 
given the super-linear rate of opinion shift given by expression (\ref{hnenthwg}), 
the crash hazard rate $h(p)$ exhibits a power law divergence of the form (\ref{rwthryt}) with
\begin{equation}
\label{eq:htsfefheo}
\alpha =
\begin{cases}
 \frac{a-\mu}{\sigma}  ~~~ ({\rm homogenous ~superlinearity~ with} ~a>\mu) ~ ({\rm expression}~
 (\ref{wrynrtbq})~{\rm with}~ (\ref{eq:sstar}))  \\
 \mu  ~~~~~~ ({\rm heterogeneous ~superlinearity~in~the~limit}~(\ref{tumiujethw}))~ ({\rm expression}~(\ref{wrthwhyfdqfdju}))
 \end{cases}
\end{equation}
}

\subsection{Relationship between the crash hazard rate and percolation susceptibility}

This result provides a micro-foundation for one of the fundamental assumptions of the JLS model, namely that the crash 
hazard rate grows and diverges as a power law singularity when the end of the bubble is approached.
The key ingredient for this behavior is the super-linear dependence (\ref{hnenthwg}) of the activity rate $R(s)$ 
that enters into the crash hazard rate expression (\ref{eq:crashhaza}). 
If $R(s) \sim s$ ($a=1$), then expression (\ref{eq:crashhaza}) leads
to a finite number since $\sum \limits_{s=1}^\infty s  ~n(s) =1$ by definition, as mentioned earlier, which means 
that the crash hazard rate is a constant.
The case where $R(s) \sim s^2$ ($a=2$) leads to $h(t)$ being proportional to the mean cluster size
defined as $\sum \limits_{s=1}^\infty s^2 ~n(s)$, which is known to diverge as $|p-p_c|^{-\gamma}$,
with $\gamma = (2-\mu)/\sigma$ (for homogenous superlinearity, the exponent $\xi$ is 
equal to $\gamma$ for $a=2$). The mean cluster size in percolation plays the role of the
susceptibility variable in other systems exhibiting a critical phase transition \cite{Stauff}.
The susceptibility quantifies how much does the order parameter respond to 
the presence of an external field. This remark makes the crash hazard rate, via its
expression (\ref{eq:crashhaza}), a kind of generalized susceptibility (for arbitrary $a>1$) to tiny external perturbations.
This provides a pleasant analogy between the usual meaning of susceptibility
and the crash hazard rate that quantifies the propensity for the price to crash, i.e. to react
to small triggering factors. As the critical point at $t_c$ is approached, this susceptibility diverges
and the crash is the most probable at this time.

\subsection{Numerical illustration with two-dimensional site percolation}

Consider traders populating a two-dimensional Manhattan-like square lattice. Traders sit on the nodes
of this lattice. When two neighbouring nodes are occupied, each by one trader, the two traders
are linked in a cluster, according to the standard definition of site percolation. As proposed above, 
each cluster of traders (set of occupied connected nodes) shares the same 
opinion (buy or sell)
on the market. Assuming that the fraction $p$ of occupied nodes increases linearly with time $t$,
the dynamics of $h(t)$ given by expression (\ref{eq:crashhaza})
is then the same as the dependence $h(p)$ as a function of $p$.
For each value of $p$, we generate a random configuration of node occupancy in a $L \times L$ lattice
and, given the specific set of obtained clusters, we evaluate expression (\ref{eq:crashhaza}) for that
particular realisation. We assume for simplicity that each value of $p$ corresponds to a reshuffled 
configuration of nodes with the corresponding occupancy fraction. This amounts to treating
each $p$ value independently. This corresponds to the limit where the traders change their
network of acquaintance faster than they may change their opinion. While this limit is 
not realistic, it actually captures the main power law dynamics of the crash hazard rate, adding
statistical fluctuations that may embody other sources of uncertainty.

\begin{figure}
\subfloat[Crash hazard rate $h_{\textrm{exp}}(p)$ (blue stars) obtained for one realisation of a 2D square lattice system of size $L=10$ for each value of $p$. The realisations are drawn independently for each value of $p$. The continuous red line is the theoretical expression $h_{\textrm{exp}}(p)$ given by formula (\ref{wrynrtbq}).]{\includegraphics[width=0.49\textwidth]{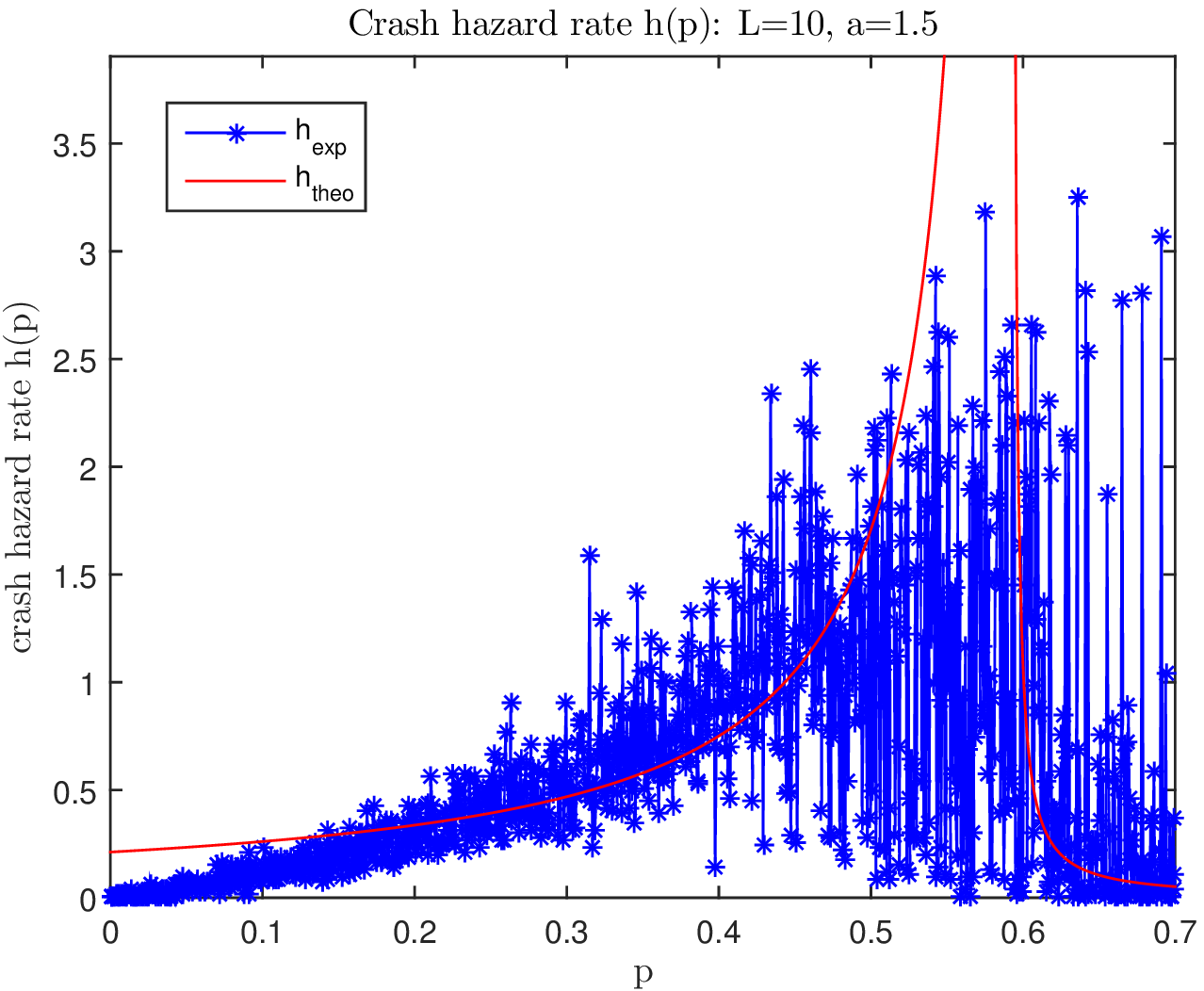}}
\hfill
\subfloat[Same as the left panel for a larger system $L=100$. Note the noticeable self-averaging observed for values of $p$
not too close to $p_c$, expressed by the much smoother $h(p)$ dependence than for the smaller system on the left panel.
Close to $p_c$, one can still observe large fluctuations as $p \to p_c$, reflecting the divergence of the susceptibility
and the sensitivity to specific realisation of clusters of sizes comparable to the system size.]{\includegraphics[width=0.49\textwidth]{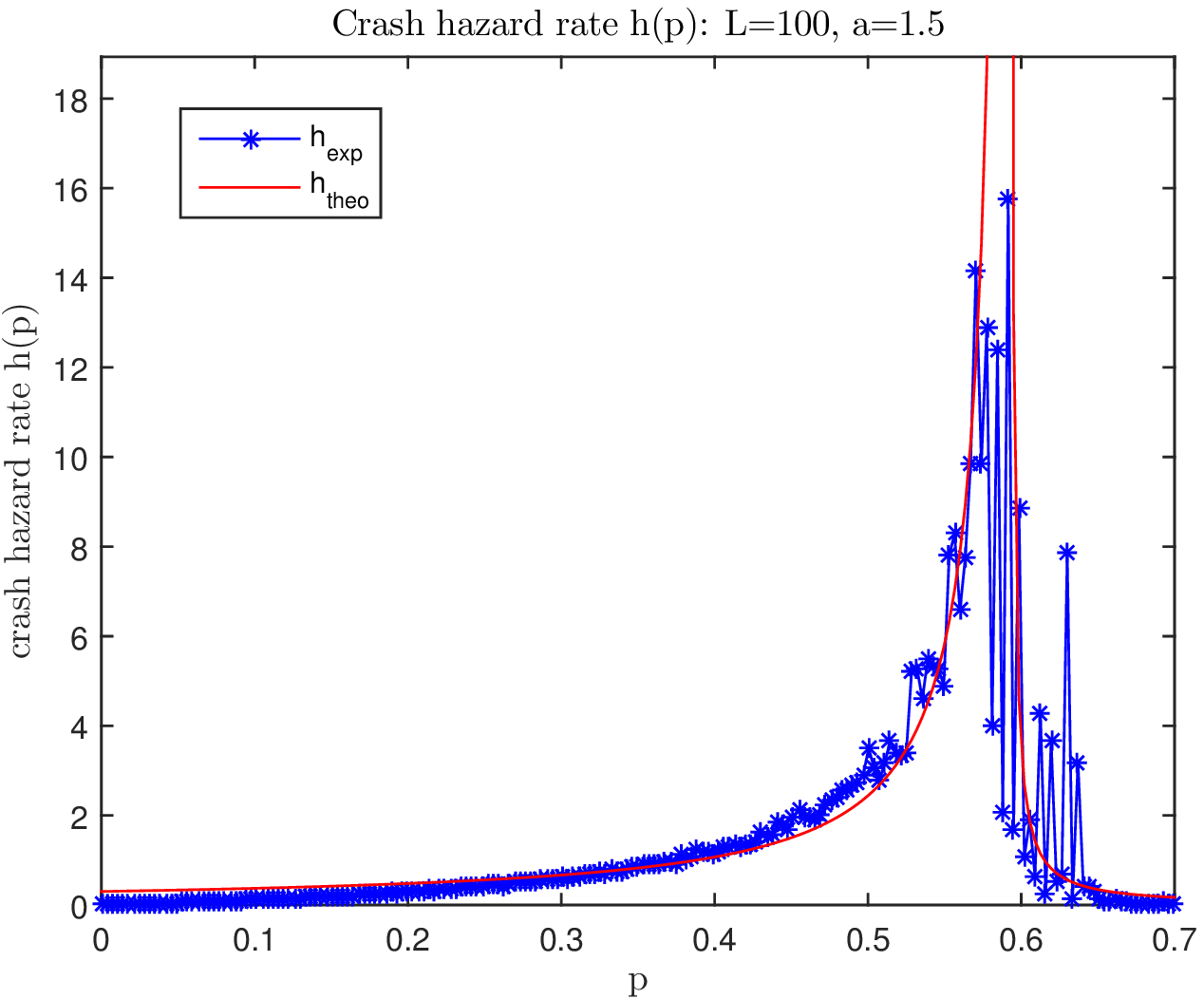}}
\caption{Crash hazard rate $h(p)$ obtained by explicit evaluation of expression (\ref{eq:crashhaza}) with a super-linear
exponent $a=1.5$ defined in (\ref{hnenthwg})
on configurations of site percolation in two dimensional square lattices.}
\label{fig:L100}
\end{figure}

Figure \ref{fig:L100} shows $h(p)$ for one realisation of a system of size $L=10$ 
(corresponding to a maximum of $100$ traders) and one
realisation for a larger system $L=100$ (maximum number of traders equal to $10^4$).
Given the specific realisation and independence of the node positions for each successive $p$'s,
it is not surprising to observe a large variance of $h(p)$. Nevertheless, the main 
prediction of an accelerated growth of $h(p)$ as $p \to p_c$ is clearly visible, even for the smallest
system size, and very vivid for the largest one. The fact that $h(t)$ fluctuates more wildly 
as $p$ increases and the amplitude of its fluctuations peak around $p_c$ results from the
appearance of larger (and thus fewer) clusters, whose specific sizes depend very sensitively 
on the configuration, the more so, the closer to $p_c$. This is nothing but the visual embodiment
of the divergence of the susceptibility of the system close to the percolation threshold.

\begin{figure}
{\includegraphics[width=0.49\textwidth]{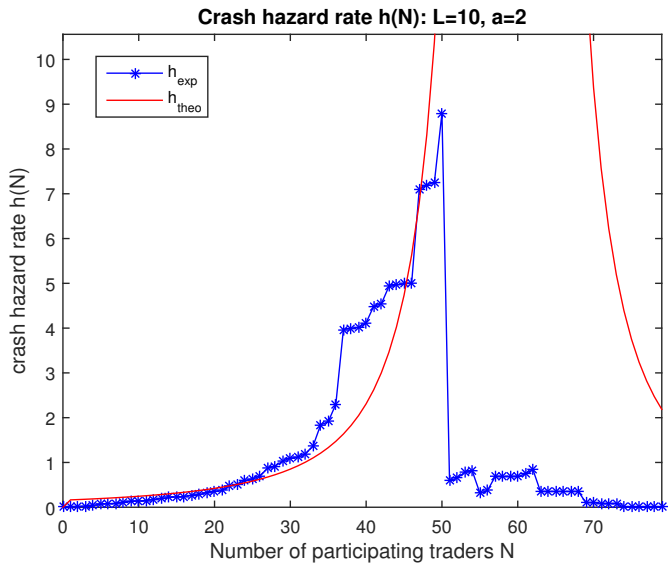}}
\caption{Same as left panel of figure \ref{fig:L100} with a super-linear
exponent $a=2$ defined in (\ref{hnenthwg}) except for the way 
the cluster configurations are constructed: starting with an empty lattice of size $L=10$, we add
traders randomly one-by-one on unoccupied sites. As $p$ increases by steps of $0.01$,
this corresponds to a system of developing 
clusters that keep the memory of previous cluster configurations at smaller $p$ values. 
The continuous red line is the theoretical expression $h_{\textrm{exp}}(p)$ given by formula (\ref{wrynrtbq}).}
\label{fig:conservative}
\end{figure}

Figure \ref{fig:conservative} is similar to the left panel of figure \ref{fig:L100}, except for the way 
the configurations are constructed. Rather than assuming that each value of $p$ corresponds to a reshuffled 
configuration of nodes with the corresponding occupancy fraction, we start from an empty lattice of size $L=10$
and put traders randomly one-by-one on unoccupied sites. The occupancy parameter $p$ thus 
varies from $0$ to $1$ in steps of $0.01$ for $L=10$. This construction assumes the other extreme
of traders fixed in their acquaintance network, only adding new relationships but never deleting any.
Unsurprisingly, the resulting crash hazard rate $h(p)$ is much smoother, while exhibiting the same
approximate power law acceleration as in the left panel of figure \ref{fig:L100}.

Similar behaviors as shown in figures \ref{fig:L100} and \ref{fig:conservative} are observed
when using the Ising model \cite{ising} instead of the percolation model. Each site
is now occupied by a spin that represents a trader, which can take one of two opinions (up or down, 
buy or sell). Using the Coniglio-Klein correspondence \cite{ConKlein1,Conklein2} and 
the Glauber dynamics \cite{Glauber}, we obtain (not shown) a dependence of the crash hazard rate 
as a function of the normalized Ising coupling constant (here playing the role of the control
parameter varying linearly with time) that is statistically identical to that shown 
in figure \ref{fig:L100}.

\section{Synthetic price Time Series}

\subsection{Implications of the finite-time singular crash hazard rate (\ref{eq:htsfefheo}) on price dynamics}

The dynamics (\ref{eq:pric2t4e}) (resp. (\ref{eq:pricesim}) in discrete time) of the price shows that,
conditional on the fact that no crash occurs, the deterministic component of the price
grows as
\begin{equation}
\ln p(t) \sim \int^t  h(\tau) ~d\tau \sim {1 \over (t_c - t)^{\alpha -1}}~, ~~~~ {\rm if} ~\alpha >1~.
\label{tyfgwc}
\end{equation}
The condition $\alpha >1$ holds in the
heterogeneous super-linearity case and in the limit (\ref{tumiujethw}) since $\alpha =\mu$,
and since the exponent $\mu$ given by (\ref{wytjuehw}) is strictly larger than $1$.
This behavior (\ref{tyfgwc}) embodies a very violent transient super-exponential growth of the price
during a bubble. 

For $\alpha <1$, as can occur when $\mu < a < \mu+\sigma$ in the homogenous super-linearity case for which 
$\alpha = (a-\mu)/\sigma$, expression (\ref{tyfgwc}) is changed into
\begin{equation}
\ln p(t) \sim \int^t  h(\tau) ~d\tau \sim  A - B (t_c - t)^{m}~, ~~~~ {\rm if} ~\alpha <1~, ~~ m :=1-\alpha~,
\label{tyfgggwc}
\end{equation}
where $A$ and $B$ are two positive constants.
This expression (\ref{tyfgggwc}) again captures a finite-time singularity at time $t_c$,
associated with a diverging local growth rate $dp/p$, while the price remains finite at $t_c$
since $0 < m < 1$.
The form is the basis of the parametric model that is used in many empirical calibrations of financial
bubbles  \cite{JLS,JLS2,Johansen1999,vladimir,ZhouSorfund06,SorZhoufor06,YanRedaetal12,Yanwoodsor12,Sorclarif13,LinSor13,YanWoodSorinf14,LinRenSor14}. Here, it stems from the rational expectation condition that links
the instantaneous return to the crash hazard rate, together with the divergence of the latter.

\subsection{Illustration by numerical simulations using a piecewise linear occupancy fraction $p=p_0 + C \cdot t$}

In this first cartoon implementation, we use a two-dimensional square lattice of size $L=500$ and
we assume that the number of traders participating in the market is linearly 
increasing with $t$ until a crash occurs, according to 
\begin{equation}
p(t)= p_0 + C \cdot t~. 
\label{ryhthbhgv}
\end{equation}
We fix $p_0=0.4$ 
and $C=10^{-4}$. For each $p(t)$, we enumerate the structure of the traders' network in terms of its cluster sizes,
which allows us to obtain the instantaneous crash hazard rate according to formula (\ref{eq:crashhaza}), 
in which the minimum cluster size contributing to the crash hazard rate is set to $s_m = 0.01\cdot p \cdot L^2$. 
The price dynamics is given by (\ref{eq:pricesim}) in which we set $\eta=0$ (no volatility). This implies the dynamics
\begin{equation}
\label{eq:pricesimwt2t}
p_{\textrm{rice}}(t)= p_{\textrm{rice}}(t-1) \left[\kappa h(t) \Delta t - \kappa \Delta j(t) \right]~,
\end{equation}
where $\Delta j(t)=0$ as long as a crash does not occur and $\Delta j(t)=1$ when a crash happens. 
Note that the crash hazard rate is nothing but the
intensity of the non-homogenous Poisson process governing the occurrence of crashes \cite{ogata}.
Poisson processes are used to simulate radioactive decay in physics \cite{raddecay}, phone calls \cite{phonecalls} in socio-science or plate tectonics spontaneous stress release in geology \cite{platetetonics}. The numerical determination of the occurrence
of a crash is made using the standard ``thinning'' algorithm for non-homogenous Poisson processes \cite{lewisshedler,Sigman}.
In our simulations, we fix the amplitude of the crash to $\kappa = 0.2$, corresponding to an instantaneous drop of 
the price of $20\%$. When a crash occurs, we also reset $p(t)$ back to $p_0=0.4$, which captures the 
fragmentation of the network of traders resulting from the bursting of the bubble observed empirically indirectly \cite{SorJohBou96}.
  
\begin{figure}
\includegraphics[width=\textwidth]{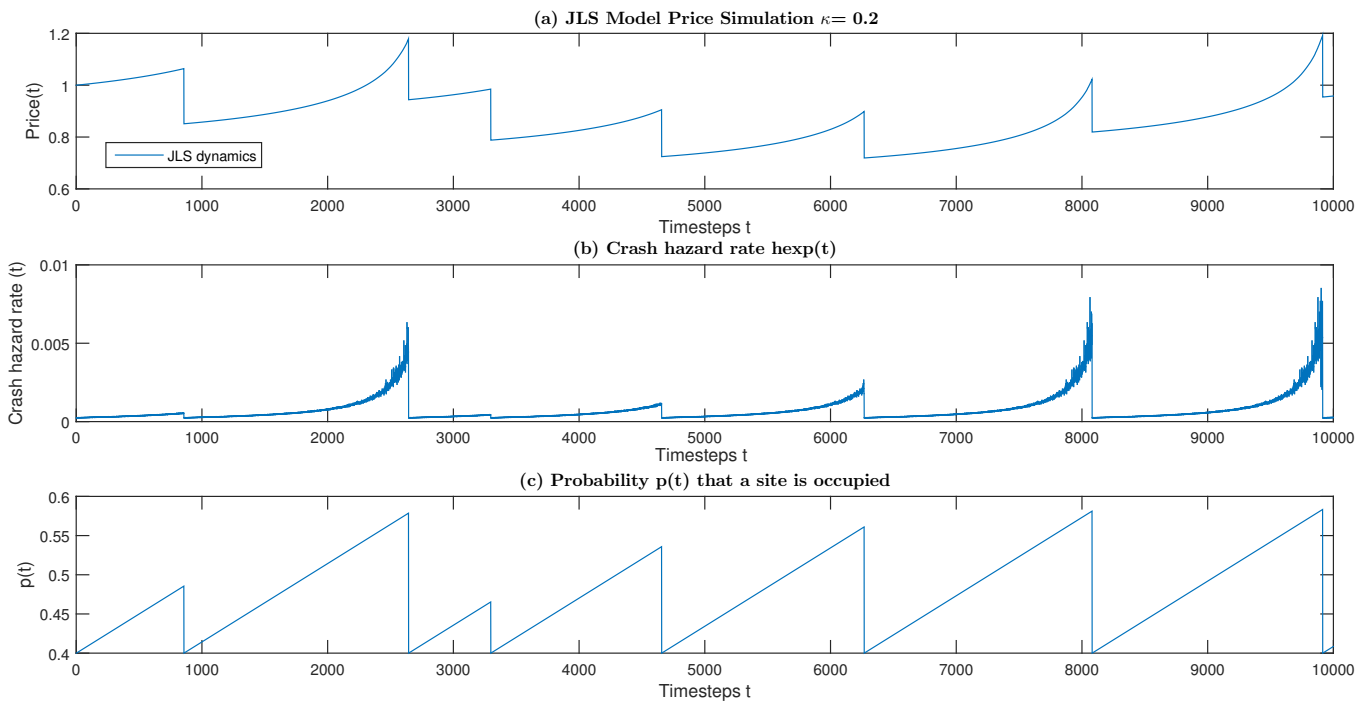}
\caption{Plot of (a) the JLS model price simulation without volatility, (b) the associated crash hazard rate $h(t)$ and (c) the linearly increasing control parameter $p(t)$ determining the crash hazard rate on a random site percolation square lattice of size $L=500$. The super-linear exponent for opinion shifts of clusters as a function of size (\ref{hnenthwg}) is set to $a=2$. The minimum cluster 
size contributing to the crash hazard rate is set to $s_m = 0.01\cdot p \cdot L^2$. Note that $p(t)$ never passes
over $p_c=0.5927...$ (for 2D site-percolation), since $h(t)$ grows very large and the crash occurs before $p_c$ is reached, resetting $p(t)$ for the next bubble.
}
\label{fig:pricelin}
\end{figure}

A typical simulation is shown in figure \ref{fig:pricelin}, where the price $p_{\textrm{rice}}(t)$, the crash hazard rate $h(t)$
and the control parameter $p(t)$ are shown over $10^4$ time steps (the unit of time is $\Delta t =10^{-4}$).
A series of bubbles can be seen, which are interrupted at various stages of their development by their corresponding crashes.
The bubbles exhibit their characteristic super-exponential trajectories, resulting from the acceleration of the
crash hazard rate as $p(t)$ increases linearly with time.
The random crash occurrences stem from their stochastic Poisson nature: while a crash is most probable
when $h(t)$ is largest, i.e., when $p(c)=p_c$ (up to finite-size effects), it can still occur at any time, as one can see
in figure \ref{fig:pricelin}. As a consequence, notwithstanding the deterministic evolution (\ref{ryhthbhgv}) 
of the control parameter $p(t)$, the crashes are non-periodic. Some randomness is also due to the stochastic
realisations of the clusters in the percolation network.
The overall resulting dynamics is stationary at long times since it is easy to check from (\ref{eq:pricesimwt2t})
that ${\rm E}[p_{\textrm{rice}}(t)] = {\rm E}[p_{\textrm{rice}}(t-1)]$,
which results from the fact that $h(t) \Delta t = {\rm E}[\Delta j(t)]$ by definition,
as explained in the above sub-section providing a summary of the mathematical formulation of the JLS model.

\subsection{Illustrations by numerical simulations using a Ornstein-Uhlenbeck $p(t)$}

In agent-based models constructed on generalisations of the Ising model \cite{HarrasTessSor12,LeissNaxSor15},
it was proposed that a natural dynamics of the control parameter $p(t)$ is not deterministic but random with persistence
in order to represent the overall social sentiment of the market. The simplest process that conveniently 
captures this is the Ornstein-Uhlenbeck process \cite{OU}, which is a mean-reverting Wiener process (or
mean-reverting random walk in discrete time). In continuous time, this reads
\begin{equation}
dp_t= \vartheta (p_0 - p_t) dt + \sigma_p dW_t~,
\label{wrthyh}
\end{equation}
where $W_t$ is a standard Brownian motion on $t \in [0,\infty)$, 
$\vartheta > 0$ is the rate of mean reversion, $p_0$ is the long-term mean of the process
and $\sigma_p >0$ is the standard deviation of the random walk excursions away from $p_0$.
 
\begin{figure}
\includegraphics[width=\textwidth]{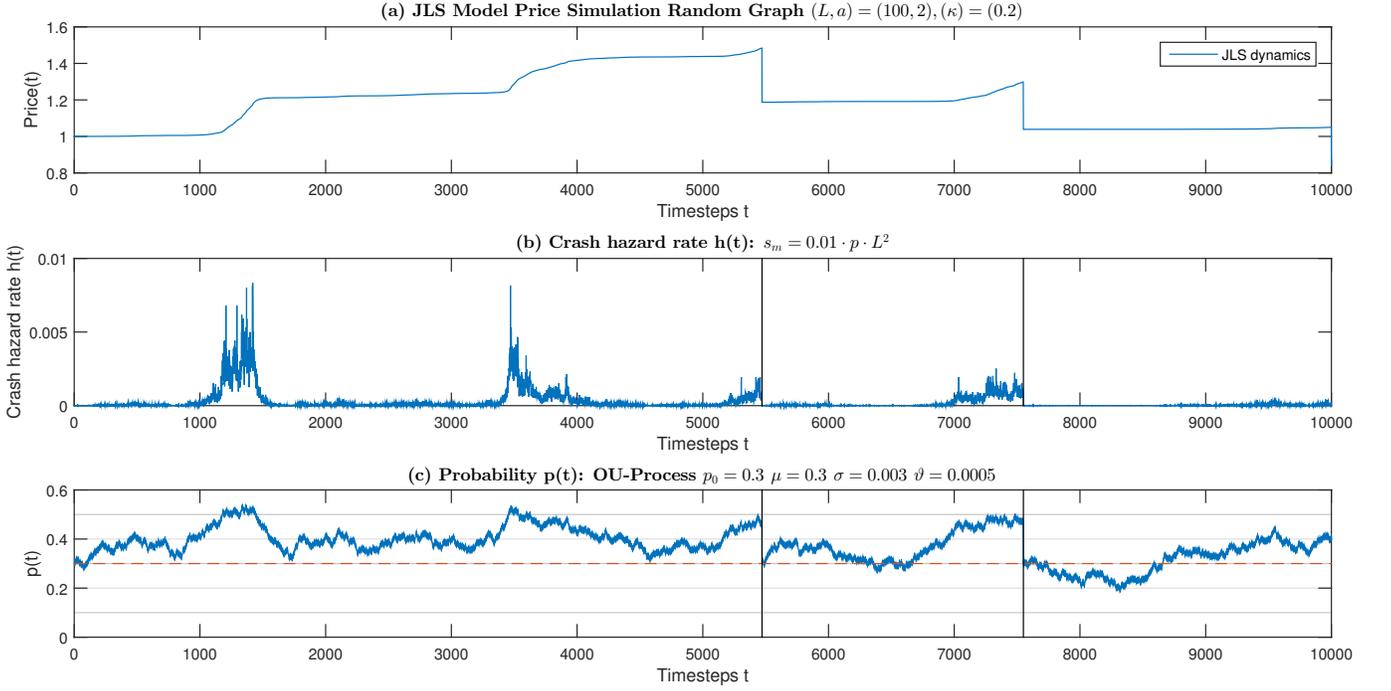}
\caption{Plot of (a) the JLS model price simulation, (b) the associated crash hazard rate and (c) the control parameter $p$ 
following an Ornstein-Uhlenbeck process, which determines the crash hazard rate on a random site percolation square lattice of size $L=100$. The values of the parameters are $a=2$, $\kappa =0.2$,  $\sigma_p=0.003$ and $\vartheta = 0.0005$. The minimum cluster size contributing to the crash hazard rate is set to $s_m = 0.01\cdot p \cdot L^2$.}
\label{fig:bubbles}
\end{figure}

With this modification (\ref{wrthyh}), we use exactly the same numerical procedure as in the previous section
to calculate the crash hazard rate and generate the price dynamics.
The only other difference is the value $p_0=0.3$ that is both
the long-term value of $p(t)$ in (\ref{wrthyh}) and the reset value after a crash.
A typical simulation is shown in figure \ref{fig:bubbles},
obtained with $L=100$, $a=2$, $\kappa =0.2$, $\sigma_p=0.003$ and $\vartheta = 0.0005$.
The value of $\vartheta$ corresponds to a memory of the Ornstein-Uhlenbeck process over typically
 $1/\vartheta = 2000$ time steps. The minimum cluster size contributing to the crash hazard rate is 
 again set to $s_m = 0.01\cdot p \cdot L^2$. 
 
Panel (b) of figure \ref{fig:bubbles} shows four periods in which the crash hazard rate increases explosively.
These periods corresponds to the four excursions of the control parameter $p(t)$ closest to $p_c=0.5927...$ (for 2D site-percolation).
Remarkably, the first two bubbles are not followed by a crash, because $p(t)$ spontaneously reverts to the mean
before a crash controlled by the conditional Poisson process with intensity $h(t)$ has time to occur.
This illustrates vividly one of the main concepts of rational expectation bubbles, on which the JLS model is build, 
namely that the end of the bubble characterized by a peak of the crash hazard rate, while being the most probable time for a crash,
is not necessarily associated with a crash. For a bubble to develop, there needs to be a finite probability 
for the bubble to end without a crash, so that rational investors can remain in the market.
Only the last two periods of large crash hazard rates end with a crash.
This is the main difference between the Ornstein-Uhlenbeck dynamics of $p(t)$ and the previous
deterministic linear increase (\ref{ryhthbhgv}), which makes the former more acceptable within 
the framework of rational expectation bubbles. Otherwise, the same quantitative properties hold.
 
\begin{figure}
\includegraphics[width=\textwidth]{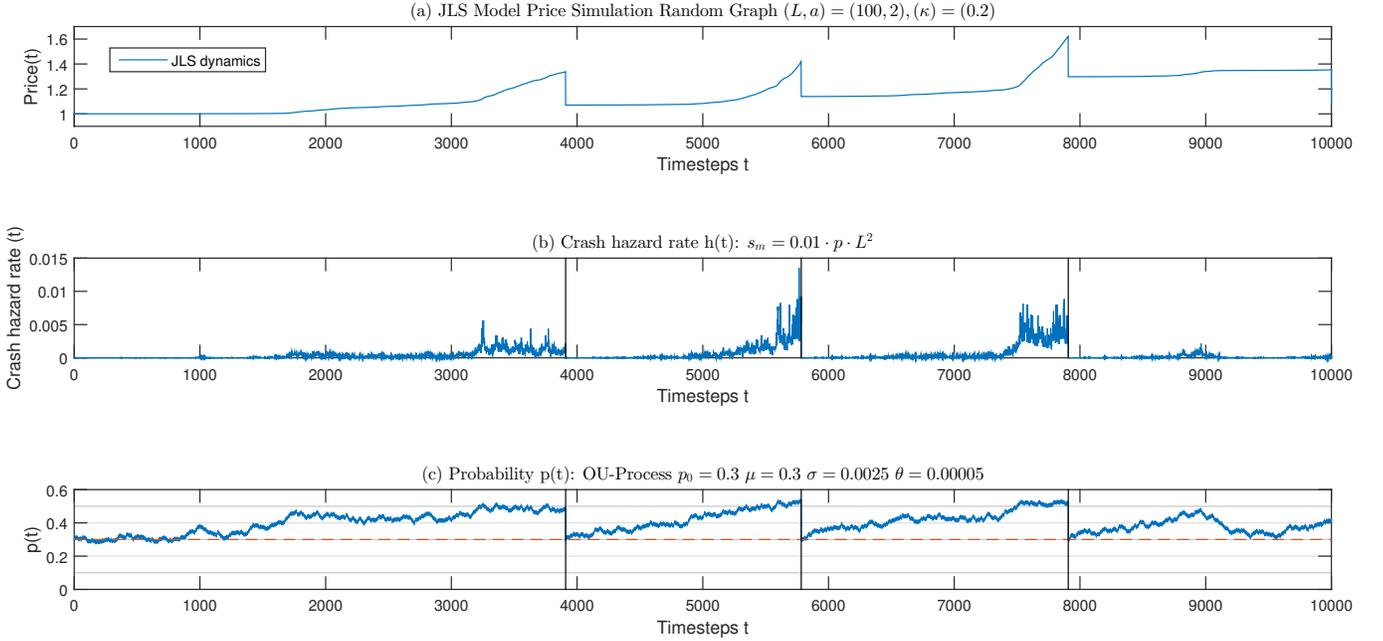}
\caption{Same as figure \ref{fig:bubbles}, except for $\vartheta = 0.00005$ corresponding to a tenfold 
increase in the memory of the Ornstein-Uhlenbeck process over typically $1/\vartheta = 20000$ time steps.
}
\label{fig:LONGMEMORY}
\end{figure}

Figure \ref{fig:LONGMEMORY} differs from  figure \ref{fig:bubbles} only with respect to $\vartheta = 0.00005$,
taken ten times smaller such that the memory $1/\vartheta$ of the Ornstein-Uhlenbeck process lasts
over $20000$ time steps, i.e., longer than the timespan shown in the figure. The situation is quite reminiscent
of what is observed for the case of the deterministic linear increase of the control parameter shown in figure \ref{fig:pricelin}.

\subsection{Illustrations by numerical simulations using a Ornstein-Uhlenbeck $p(t)$ and volatility}

\begin{figure}
\includegraphics[width=\textwidth]{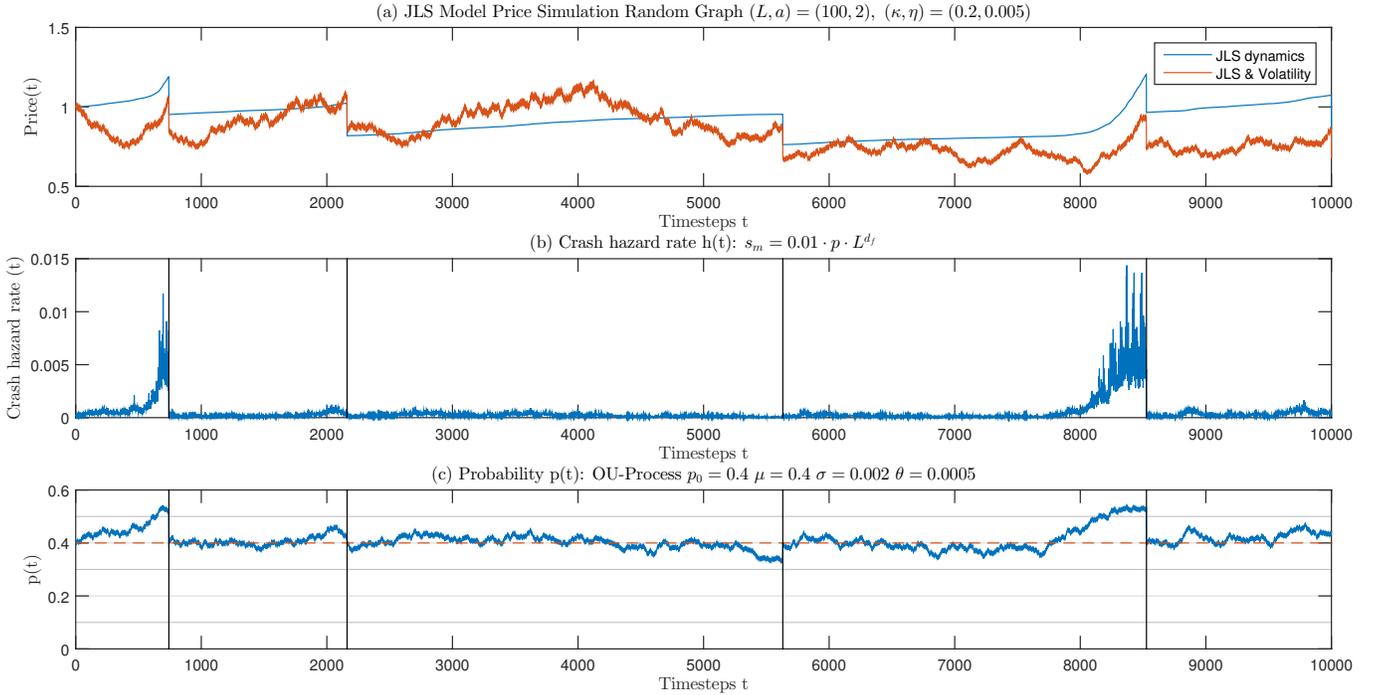}
\caption{Plot of (a) the JLS model price simulation with (red) and without (blue) the volatility term, (b) the associated crash hazard rate and (c)  the control parameter $p$  following an Ornstein-Uhlenbeck process, which determines the crash hazard rate on a random site percolation square lattice of size $L=100$.  The parameters are $a=2, \kappa=0.2, p_0=0.4, \sigma_p=0.002, \vartheta = 0.0005$ and   
the volatility of the price is set to $\eta = 0.005$. The value of the minimum cluster size contributing to the crash hazard rate is taken equal to $s_m=0.01 \cdot  p \cdot L^{d_f} (\approx 30)$, where $d_f = d - {\beta \over \nu} = 91/48$ in two dimensions.}
\label{fig:CORRinteresting}
\end{figure}

We add the last ingredient of model (\ref{eq:pricesim}), which is the volatility term $\eta \Delta w(t)$.
Figure \ref{fig:CORRinteresting} presents a typical simulation obtained with the parameters
$L=100, a=2, \kappa=0.2, \eta = 0.005, p_0=0.4, \sigma_p=0.002, \vartheta = 0.0005$ and 
the value of the minimum cluster size contributing to the crash hazard rate is taken equal to $s_m=0.01 \cdot  p \cdot L^{d_f} (\approx 30)$,
where $d_f = d - {\beta \over \nu} = 91/48$ in two dimensions is the fractal dimension of percolation clusters.
While secondary for our present simulation made for a relatively small system $L=100$, 
this choice of the scaling $s_m \sim  p \cdot L^{d_f}$ will ensure
that the limit of very large system sizes $L \to +\infty$ can be taken while keeping the results non trivial. 
Indeed, $L^{d_f}$ is the typical number of occupied sites at the percolation threshold $p_c$, so that the choice
$s_m \sim  p \cdot L^{d_f}$ ensures that only the ``large'' clusters in an asymptotic sense contribute
to destabilizing the market to create crashes.

Figure \ref{fig:CORRinteresting} exhibits four crashes. The first and last ones are associated with periods
of explosive growth of the crash hazard rate, and thus of the price that presents its characteristic super-exponential
shape. These two periods coincide with the excursions of the control parameter $p(t)$ close to $p_c$.
The second crash occurs at a time when the crash hazard is larger than most of the time but still quite small
compared to the two explosive periods. The third crash occurs at a time of an uneventful dynamics of $h(t)$.
This can indeed occur, given the stochastic nature of the Poisson process triggering the crash, even when 
the corresponding intensity is small (but non zero). 

This typical simulation shows that the interplay between the Ornstein-Uhlenbeck dynamics of the
control parameter $p(t)$, the construction of the crash hazard rate with explosive behavior and its
impact on the price via the rational expectation condition together with the volatility component
create realistic looking price processes.

\section{Concluding remarks}
 
The purpose of the present article has been to present a plausible micro-founded model for the previously postulated
power law finite time singular form of the crash hazard rate in the JLS model \cite{JLS,JLS2,Johansen1999}.
The model is based on a percolation picture of the network of traders and the concept
that clusters of connected traders share the same opinion. The key ingredient is the notion that a shift
of position from buyer to seller of a sufficiently large group of traders can trigger a crash.
This provides a formula to estimate the crash hazard rate by a summation over percolation clusters
above a minimum size of a power of the cluster sizes, similarly to a generalized percolation susceptibility.
The power of cluster sizes emerges from the super-linear dependence of group activity as a function
of group size, previously documented in the literature. The crash hazard rate was shown to exhibit
explosive finite-time singular behavior when the control parameter (fraction of occupied sites, or density
of traders in the network) approaches the percolation threshold $p_c$. 
Then, within the framework of rational expectation bubbles, the instantaneous return must
be proportional to the crash hazard rate in order to compensate investors for taking the risk of crashes.
During the bubbles preceding the crashes, the price then grows super-exponentially as a consequence.

Our theoretical derivations and numerical investigations, of which only a few are
shown in the above figures, can be summarized as follows.

\begin{enumerate}
\item The JLS model exhibits intermittent explosive super-exponential bubbles interrupted by crashes. 
Their combination ensures a long-term diffusive behavior without drift for the log-price.
\item Bubbles are generated by the excursion of the control parameter $p(t)$, quantified by the fraction 
of occupied sites on a percolation network, close to the percolation threshold $p_c$.
\item The memory of the Ornstein-Uhlenbeck process controlling the dynamics of the control parameter $p(t)$
modulates the frequency of the bubbles via the frequency with which $p(t)$ deviates away and reverts towards
its long-term average.
\item Bubbles can end sometimes without a crash, when the control parameter $p(t)$ spontaneously reverts to the mean
before a crash controlled by the conditional Poisson process with intensity $h(t)$ has time to occur.
\item During bubbles, the price grows super-exponentially (with upward convexity of the logarithm of the price
as a function of time), due to the proportionality
between the instantaneous return and explosive crash hazard rate.
\item The higher the super-linear exponent $a$ for the change of opinion of clusters, the sharper and larger the bubbles can get. Crashes occur more frequently.
\item A longer memory of the Ornstein-Uhlenbeck process leads to larger bubbles that end very likely with a crash. A shorter memory of the 
Ornstein-Uhlenbeck process leads to small bubbles that occur very often, but more rarely end with a crash.
\item The larger the price volatility $\eta$, the more likely it is that it may amplify or hide the development of a bubble.
\end{enumerate}

\bibliography{JLSMicrofoundationSeyrichSornette}
\bibliographystyle{unsrt}

\end{document}